\documentclass[aps,prc,twocolumn,groupedaddress,showpacs,superscriptaddress,floatfix,preprintnumbers]{revtex4-2}

\usepackage{amsmath}\allowdisplaybreaks[1]
\usepackage{graphicx}
\usepackage{amssymb}
\usepackage{bm}
\usepackage{array}
\usepackage{xcolor}
\usepackage[colorlinks,linkcolor=blue,urlcolor=blue,anchorcolor=blue,citecolor=blue]{hyperref}
\usepackage{slashed}
\usepackage{braket}
\usepackage{tabularx}
\graphicspath{{./fig/}}

\begin{document}

\title{Subthreshold production of $J/\psi$ mesons from the deuteron with SoLID}

\newcommand*{\SDU}{Key Laboratory of Particle Physics and Particle Irradiation (MOE), Institute of Frontier and Interdisciplinary Science, Shandong University, Qingdao, Shandong 266237, China}\affiliation{\SDU}
\newcommand*{\SCNT}{Southern Center for Nuclear-Science Theory (SCNT), Institute of Modern Physics, Chinese Academy of Sciences, Huizhou 516000, China}\affiliation{\SCNT}
\newcommand*{\DUKE}{Department of Physics, Duke University and Triangle Universities Nuclear Laboratory, Durham, North Carolina, USA}\affiliation{\DUKE}
\newcommand*{\PKU}{Department of Physics, Beijing University, Beijing, China}\affiliation{\PKU}
\newcommand*{\ITP}{CAS Key Laboratory of Theoretical Physics, Institute of Theoretical Physics, Chinese Academy of Sciences, Beijing 100190, China}\affiliation{\ITP}

\author{Tianbo Liu}\affiliation{\SDU}\affiliation{\SCNT}
\author{Zhiwen Zhao}\affiliation{\DUKE}
\author{Mengchu Cai}\affiliation{\PKU}\affiliation{\ITP}
\author{Duane Byer}\affiliation{\DUKE}
\author{Haiyan Gao}\affiliation{\DUKE}



\begin{abstract}
The electro- and photo-production of $J/\psi$ meson near the threshold from the proton is relevant to the search of hidden charm pentaquark candidates reported by the LHCb collaboration, and the study of the QCD trace anomaly's contribution to the proton mass. 
It is also expected to be sensitive to
the QCD van der Waals interaction, that is mediated by multi-gluon exchanges and expected to dominate the interaction between two hadrons with no common valence quarks. 
Subthreshold production of $J/\psi$ from a nuclear target is expected to enhance such attractive interaction, and also allows for a direct probe of short range correlations inside a nucleus. With the high luminosity capability of the 12-GeV CEBAF facility at Jefferson Lab, high-precision data on $J/\psi$ meson production from the proton is becoming available, providing also a reference for subthreshold $J/\psi$ production from the deuteron.  Data from the deuteron will establish the baseline for subthreshold $J/\psi$ production from other nuclear targets. In this paper, we present our findings from a feasibility study of subthreshold $J/\psi$ production from the deuteron using the proposed Solenoidal Large Intensity Device (SoLID), and discuss the potential physics impact of such data.       

\end{abstract}

\maketitle

\section{Introduction}
\label{sec:intro}
The $J/\psi$ meson has been a fascinating particle to physicists since its original discovery in the 1970s~\cite{E598:1974sol,SLAC-SP-017:1974ind}. Its discovery -- the discovery of the charm quark -- played an important role in the establishment of the quark model as a credible description of nature. In recent years, the $J/\psi$ meson is also playing an important role in the search of multi-quark states~\cite{Chen:2016qju,Liu:2019zoy}, one of the active frontiers since the establishment of the quark model~\cite{GellMann:1964nj,Zweig:1981pd}. Hadrons, as color singlet states, are composite particles made of quarks and gluons held together by strong interactions. Ordinary mesons are described as quark-antiquark states and ordinary baryons are described as three-quark states. Exotic mesons, such as glueballs, tetraquarks, and hybrid mesons, have quantum numbers that are impossible for a quark-antiquark configuration. 
Similarly, exotic baryons have constituents other than the three-quark configuration. The pentaquark is a type of exotic baryons whose minimum-quark-content is five-quark. 
The quantum chromodynamics (QCD) is known as the fundamental theory of strong interactions in the framework of the Yang-Mills gauge theory with quarks and gluons as degrees of freedom. It allows the existence of multi-quark states;  however, due to the non-perturbative nature of QCD at low energy scales, first principle calculations of hadronic states and properties remain challenging. Hence experimental investigation of multi-quark states is an important approach to understand the dynamics of the strong interaction at the hadronic scale.

In 2015, the LHCb collaboration discovered two hidden charm pentaquark candidates, named as $P_c(4380)$ and $P_c(4450)$, in the channel $\Lambda_b\to J/\psi K^-p$~\cite{Aaij:2015tga}, which contains narrow peaks in the $J/\psi p$ invariant mass distribution. In 2019, LHCb followed up with an order of magnitude more data. They discovered another narrow pentaquark candidate $P_c(4312)^+$ and confirmed that $P_c(4450)^+$ consists of two narrow overlapping peaks, $P_c(4440)^+$ and $P_c(4457)^+$~\cite{Aaij:2019vzc}. 
In light of these developments, the production of $J/\psi$ mesons near the threshold from protons with high precision has become ever more important and compelling.  The 12-GeV CEABF at Jefferson Lab (JLab) provides excellent opportunities to study the production of $J/\psi$ mesons near the threshold as demonstrated by the GlueX experiment in JLab Hall D first in 2019~\cite{Ali:2019lzf} and again in 2023~\cite{PhysRevC.108.025201} with more statistics.  
The GlueX collaboration reported on the measurement of the $\gamma p \rightarrow J/\psi p$ cross section for photon energies from 11.8\,GeV down to the threshold energy of 8.2\,GeV using a tagged photon beam. The results show that the total cross section falls toward the threshold less steeply than expected from two-gluon exchange models~\cite{Brodsky:2000zc}. While the LHCb pentaquark candidates $P_c^+$ in principle could be produced in the $s$ channel of this reaction, the GlueX results provide no evidence for these states. JLab experiment E12-16-007~\cite{E12-16-007} took data in Hall C in spring 2019 on photoproduction of $J/\psi$ on the proton from the threshold of 8.2\,GeV to energies above the threshold for the production of the $P_{c}(4450)$ state. The results~\cite{HallCjpsinature} have no signal on the $P_{c}(4450)$ state, but the proton gluonic gravitational form factors are extracted. 

Another major physics motivation for precise information on the near threshold photoproduction cross section of the $J/\psi$ meson from the proton is due to its connection to the QCD trace anomaly, which is predicted to contribute significantly to the mass of the proton. The proton, as the bound state of strong interaction, is fundamentally described in terms of quarks and gluons degrees of freedom in QCD. Thus, a decomposition of the proton mass can be investigated from the QCD energy-momentum tensor, which can be uniquely separated into a traceless part and a trace part following the procedure in Refs.~\cite{Ji:1994av,Ji:1995sv}. According to the separation of the energy-momentum tensor, one can partition the proton Hamiltonian in terms of the quark energy, gluon energy, quark mass, and the trace anomaly contributions. The parameter that determines the trace anomaly contribution to the proton mass, which also contributes to the quark mass and quark energy pieces can be extracted via the purely real amplitude of the interaction between heavy quarkonium and light hadron at low energy~\cite{Kharzeev:1995ij,Kharzeev:1998bz}.  The heavy quarkonium, such as $J/\psi$, is a strongly bound state of two heavy quarks with both the constituents' mass and the binding energy much larger than $\Lambda_{\textrm{QCD}}\sim340\,{\rm MeV}$~\cite{Deur:2014qfa}, the typical nonperturbative QCD scale. Thus, it can be utilized as a microscopic probe of the structure of light hadrons, such as the proton. The quarkonium-nucleon scattering $J/\psi N\rightarrow J/\psi N$ is related to the differential cross section of forward photoproduction $\gamma N \rightarrow J/\psi N$ process through the conventional vector meson dominance (VMD) approach~\cite{Kharzeev:1998bz}.

In the last two decades or so, there has been a renewed interest in the QCD van der Waals interaction -- mediated by multi-gluon exchanges -- which is expected to dominate the interaction between two hadrons when they have no common valence quarks. Such an attractive interaction in the case of the $J/{\psi}-N$ has been confirmed by both the effective field theory method~\cite{Peskin:1979va,Luke:1992tm,Brodsky:1997gh,Kaidalov:1992hd} and lattice QCD~\cite{Kawanai:2010ev}. This van der Waals attractive interaction also leads to the prediction of a bound state between a $J/\psi$ and a light nucleus~\cite{Brodsky:1989jd}. 
Lattice QCD simulations~\cite{Beane:2014sda} have been performed for both the charmonium and the strangeonium cases showing that the interaction between the charmonium/strangeonium and the nucleon/nucleus could be strong enough to form a bound state. The study of bound $J/\psi$ nuclear states~\cite{Wu:2012wta} from photoproduction off nuclear targets requires precise measurements of $J/\psi$ photoproduction from the proton and the deuteron targets near the threshold.
Subthreshold production of $\phi$ or $J/\psi$ mesons from a nuclear target is expected to enhance such attractive interactions. The $\phi-N$ bound state was explored by a number of authors~\cite{Gao:2000az,Gao:2017hya,Huang:2005gw,He:2018plt}. Gao, Lee, and Marinov~\cite{Gao:2000az} were the first to propose and showed that the subthreshold quasi-free $\phi$ meson production inside a nuclear medium would enhance the probability for the formation of the $\phi-N$ bound state. Experimental demonstration of subthreshold production of $\phi$ and $J/\psi$ mesons will be an important step towards searching and/or discovering aforementioned exotic QCD bound states. Previously, ``subthreshold" production of $\phi$ mesons from a deuterium target has been demonstrated by CLAS6~\cite{Qian:2010rr}, in which the physical threshold of the $\phi$ meson photo-production from the proton was determined by the detection acceptance of the decay kaon pair. 
A quasifree model is able to describe the differential cross section well~\cite{Qian:2010rr}, though the data have rather large statistical uncertainties.  
Subthreshold production of $J/\psi$ meson was also attempted previously in Hall C at the 6-GeV CEBAF~\cite{Bosted:2008mn}. While no event was seen for such production, this null finding is not surprising as subthreshold cross sections are small, and 5.7\,GeV used for the experiment is too far from the threshold of 8.2\,GeV.  

Recently Hatta {\it et al.}~\cite{Hatta:2019ocp} proposed subthreshold photoproduction of $J/\psi$ from nuclear targets as an independent test of the universality of the nucleon-nucleon short-range correlation in nuclear scattering processes, and the deuteron is presented as the reference nucleus for such a test.
Wu and Lee~\cite{Wu:2013xma} studied quasi-free production of $J/\psi$ from the nucleon inside the deuteron near the threshold, and found at forward proton scattering angles and low $J/\psi$ momenta in the center-of-momentum (c.m.) frame of the $J/\psi-N$ subsystem, the differential cross section is dominated by the $J/\psi-N$ rescattering. Therefore, carefully designed experiments using a deuterium target will allow for the extraction of the $J/\psi-N$ potential. Both sub- and near-threshold production of $J/\psi$ is advantageous in this case to access low $J/\psi$ momentum, therefore the rescattering region. 

In this paper, we present a study of subthreshold photoproduction of $J/\psi$ meson from the deuteron using the proposed solenoidal large intensity device (SoLID) in Hall A at JLab. While our focus is on the subthreshold production, we also present results near the threshold -- the comparison of the production mechanism of $J/\psi$ from the deuteron between subthreshold and above-the-threshold may uncover interesting short-range physics such as the nucleon-nucleon short-range correlation, non-nucleonic configuration such as hidden color states~\cite{Brodsky:1983vf}. Furthermore, data on subthreshold production of $J/\psi$ from the deuteron compared to those of the $\phi$ meson~\cite{Qian:2010rr} can uncover interesting QCD dynamics that might be reflected in the difference between charmonium and strangeonium production below the threshold on the proton.    
In addition to the photoproduction case, we will also present results on the quasi-real photoproduction. The rest of the paper is organized as follows. In Section II, we present an description of the SoLID and the experimental method for the subthreshold and near threshold $J/\psi$ production from a deuterium target. In Section III, we present our results from the simulations including the experimental apparatus, and 
the model used in our study on $\gamma p \rightarrow J/\psi p$ reaction. Finally in Section IV, we will conclude and provide future outlook.

\section{Experimental setup}
\label{sec:setup}

The SoLID is a major new device combining large acceptance and high luminosity proposed for Hall A in the JLab 12-GeV era. The approved SoLID programs~\cite{SoLID-reviewpaper} consist of an experiment on parity-violating deep inelastic scattering~\cite{PVDIS} to provide an unique test of the standard model and search for new physics, experiments on semi-inclusive deep inelastic scattering to perform tomography of the nucleon in three-dimensional momentum space~\cite{Gao:2010av}, and high precision measurements of cross sections from $J/\psi$ electroproduction and photoproduction near the production threshold from the proton~\cite{E12-12-006}.

\begin{figure}[htp]
\begin{center}
\includegraphics[width=\columnwidth]{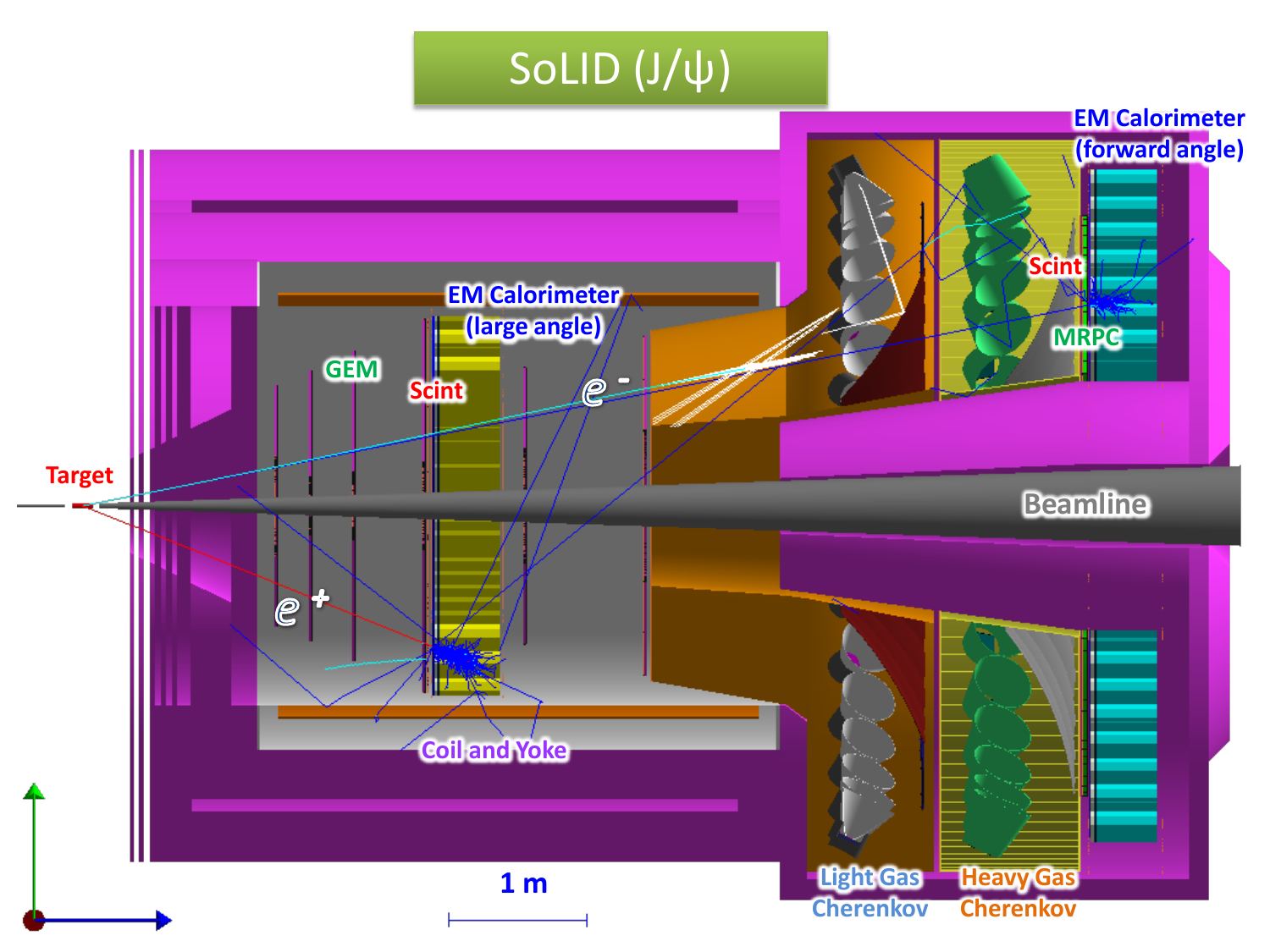}
\caption{The SoLID $J/\psi$ experimental layout.}
\label{figure:layout}
\end{center}
\end{figure}

\begin{figure}[htp]
\begin{center}
\includegraphics[width=\columnwidth]{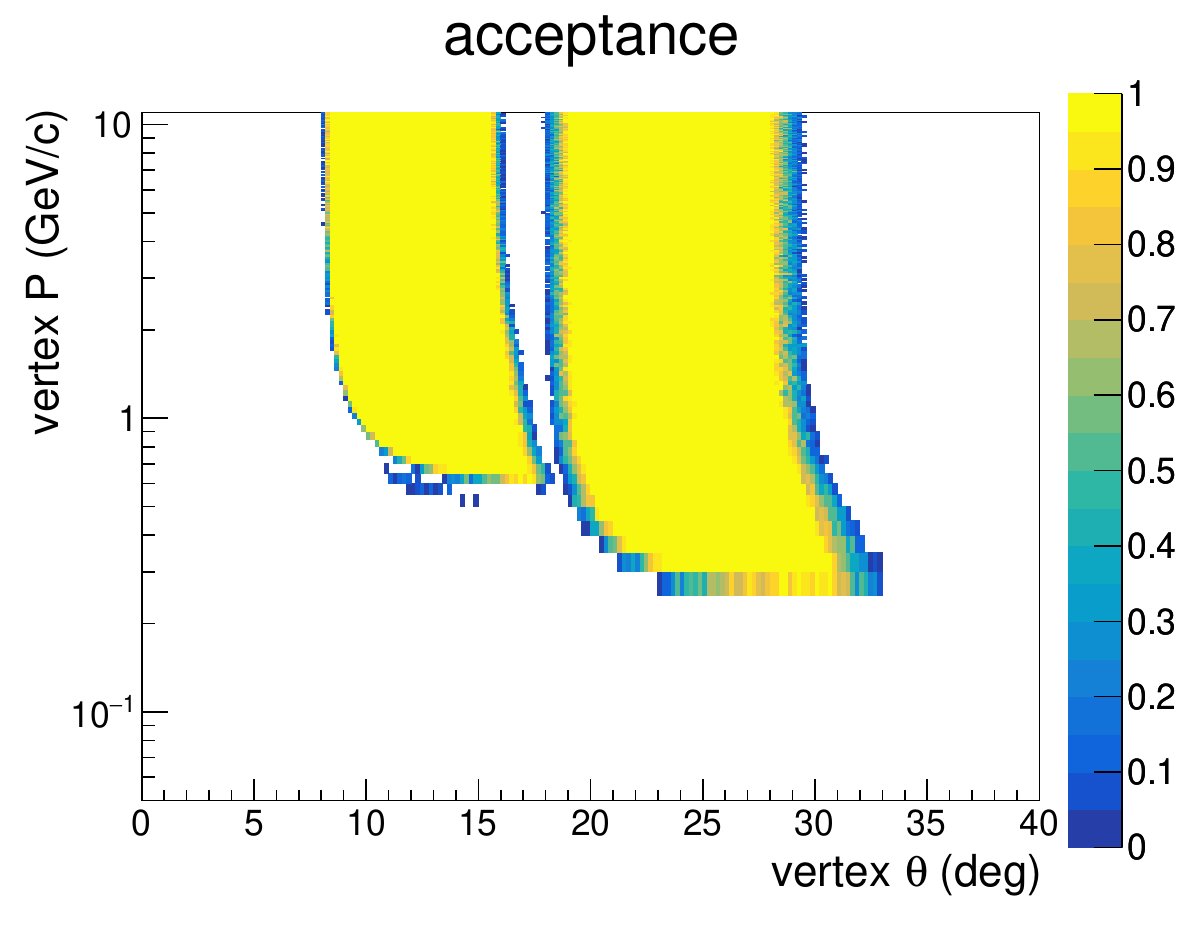}
\caption{The geometrical acceptance of a single particle in SoLID $J/\psi$ setup at the forward angle (left) and large angle (right). The momentum and polar angle are from the particle vertex when it is produced at the target.}
\label{figure:acc}
\end{center}
\end{figure}

The $J/\psi$ sub- and near-threshhold production cross sections from the deuterium can be measured using the same experimental setup as that for the approved experiment~\cite{E12-12-006} with a liquid deuterium target instead of a liquid hydrogen target.
The layout of the SoLID $J/\psi$ setup is shown in Fig.~\ref{figure:layout}. With custom designed high rate and high radiation tolerant detectors, the SoLID $J/\psi$ setup can carry out experiments using high intensity electron beams incident on a cryogenic target in an open geometry with a full azimuthal angular coverage. There are two groups of sub-detectors. The forward angle detector group covers a polar angular range from $8^\circ$ to $16^\circ$ and consists of several planes of gas electron multipliers (GEM) for tracking, a light gas Cherenkov (LGC) for $e/\pi$ separation, a heavy gas Cherenkov (HGC) for $\pi/K$ separation, a multi-gap resistive plate chamber (MRPC) for time-of-flight, a scintillator pad (SPD) for photon rejection, and a forward-angle electromagnetic calorimeter (FAEC). The large-angle detector group covers a polar angular range from $18^\circ$ to $28^\circ$ and consists of several planes of GEM for tracking, an SPD and a large-angle electromagnetic calorimeter (LAEC). Electrons, positrons, and protons will be detected and identified by measuring their momenta, time-of-flight, photons produced in the threshold Cherenkov detectors, and energy losses in the calorimeters. A single particle geometrical acceptance is shown in Fig.~\ref{figure:acc} with the target center located at 315\,cm upstream from the solenoid coil center. Limited by time-of-flight resolutions, the proton identification is good up to a momentum value of 4.5\,GeV in the forward angular region and up to 2\,GeV in the large angular region. 

Because the subthreshhold production is a rare process with small cross sections, we include both quasi-real photoproduction (low $Q^2$ electroproduction) and real photoproduction to maximize event counts. As the proton detection requires the momentum greater than 0.2\,GeV and there is no neutron detection capability in SoLID, we do not consider the production of $J/\psi$ from the neutron inside a deuteron in our study. We require a 3-fold coincident detection in which the $J/\psi$ decay pairs of electron and positron as well as recoil protons will be detected. The detection of the scattered electrons is not required even in the case of the low $Q^2$ electroproduction. 
Thus the reactions of interest are $e^- + ``p" \rightarrow (e^{-'}) + J/\psi + p \rightarrow (e^{-'}) + e^+ + e^- + p $ and $\gamma + ``p" \rightarrow J/\psi + p \rightarrow e^- + e^+ + p $,
where $J/\psi$ is reconstructed through its decay into a lepton pair ($e^{+}e^{-}$) with a branching ratio of 5.94\% and $``p"$ stands for the proton inside the deuteron.

In our study, the main trigger used is the 3-fold coincidence of the decay lepton pair and the recoil proton. There will be some 4-fold electroproduction coincident events including the scattered electrons in the obtained data set for us to study and such a topology shall provide the best signal to background ratio but low statistics. One may choose to include only 2-fold coincident detection of the $J/\psi$ decay lepton pair if the trigger design allows, but it is expected to have large backgrounds for such a topology. 

In our study, we use beam currents up to 1.25\,$\mu$A to ensure the luminosity to stay below $1.2\times10^{37}$\,${\rm nucleon}\cdot{\rm cm}^{-2}\cdot s^{-1}$, which is the same as that of the E12-12-006 experiment~\cite{E12-12-006}. The SoLID collaboration is actively working on improving the $J/\psi$ setup's capability to handle even higher luminosity. If this effort is successful, one can increase the beam current for this experiment accordingly to achieve more statistics. 

\section{Simulation}
\label{sec:simulation}

To demonstrate the experimental feasibility of the subthreshold and near-threshold $J/\psi$ production from a deuterium target, we perform a Monte Carlo simulation using the proposed SoLID detector at JLab. In this section, we will briefly introduce the theoretical model of the photoproduction of $J/\psi$ from the nucleon, and then describe the simulation procedure of the photoproduction and the electroproduction of $J/\psi$ from a deuterium target. The simulation results are presented in the end of this section.

\subsection{$J/\psi$ photoproduction model}
\label{subsec:Pom-pot}

Although there are many phenomenological models for the $J/\psi$ photoproduction from the nucleon, very few models can simultaneously describe the data at low and high energies, especially when including the precise data from JLab~\cite{Ali:2019lzf}. In a recent review~\cite{Lee:2022ymp}, the Pomeron-potential model~\cite{Lee:2020iuo}, referred to as the ``Pom-pot'' model, was found the only one that can fit data from near threshold~\cite{Ali:2019lzf} up to high energy $W = 300\,\rm GeV$~\cite{H1:1996gwv}. Hence we use the differential cross section calculated by the Pom-pot model in the simulation.

The Pom-pot model is an extension of the Pomeron-exchange model. As advanced by Donnachie and Landshoff~\cite{Donnachie:1984xq,Donnachie:1992ny,Donnachie:1994zb,Donnachie:1998gm}, the vector meson photoproduction at high energies can be well described by the Pomeron-exchange mechanism~\cite{Donnachie:2002en} and the vector meson dominance (VMD) assumption~\cite{Sakurai:1960ju,Gell-Mann:1961jim,Sakurai:1969ss,Sakurai:1972wk}. In this model, referred to as the ``Pom-DL'' model, the photon is first converted to the vector meson and then scattered by Pomeron-exchange between quarks in the vector mesion and the nucleon. Though the Pom-DL model can well fit the $J/\psi$ photoproduction data up to $W=300\,\rm GeV$, the calculated cross section at low $W$ is significantly below the recent JLab data. In such near threshold region, the relative velocity between $J/\psi$ and the nucleon is not large and higher order multi-gluon exchanges that are not included in the Pomeron-exchange mechanism cannot be neglected. To solve this problem, it was proposed in Ref.~\cite{Lee:2020iuo} to add an effective $J/\psi-N$ potential term phenomenologically interpreted as the multi-gluon exchange.

With the VMD assumption, one can calculate $J/\psi$ photoproduction cross section following the Hamiltonian,
\begin{align}
H = H_0 + V(r) + \frac{e M_{\tiny J/\psi}^2}{f_{\tiny J/\psi}} \int {\rm d}^4 x A_{\mu}(x) \phi^{\mu}(x),
\end{align}
where $f_{\tiny J/\psi}=11.2$ is the VMD coupling, which is evaluated from the decay width of $J/\psi\to e^+ e^-$. $A_\mu(x)$ and $\phi^\mu(x)$ are the field operators of the photon and the $J/\psi$ respectively, $e$ is the electric charge of the positron, and $M_{\tiny J/\psi}$ is the mass of the $J/\psi$ meson. $V(r)$ is the $J/\psi-N$ potential, which can be extracted from lattice QCD calculation~\cite{Kawanai:2010ev}. As found in Ref.~\cite{Lee:2020iuo}, the lattice QCD data can be fitted by a Yukawa form potential,
\begin{align}
V(r) &= V_0 \frac{e^{-a r}}{r},
\end{align}
where $V_0$ and $a$ are free parameters. Two sets of parameters, $V_0 = -0.06(-0.11)$ and $a = 0.3(0.5)\,\rm GeV$, are provided in Ref.~\cite{Lee:2020iuo} corresponding to two different selected fit regions of the lattice data, but the final numerical results are almost indistinguishable. More detailed discussions on the fit of the potential and the dependence on the parametrization form can be found in Ref.~\cite{Lee:2020iuo}.

\begin{figure}[htp]
\centering
\includegraphics[width=\columnwidth]{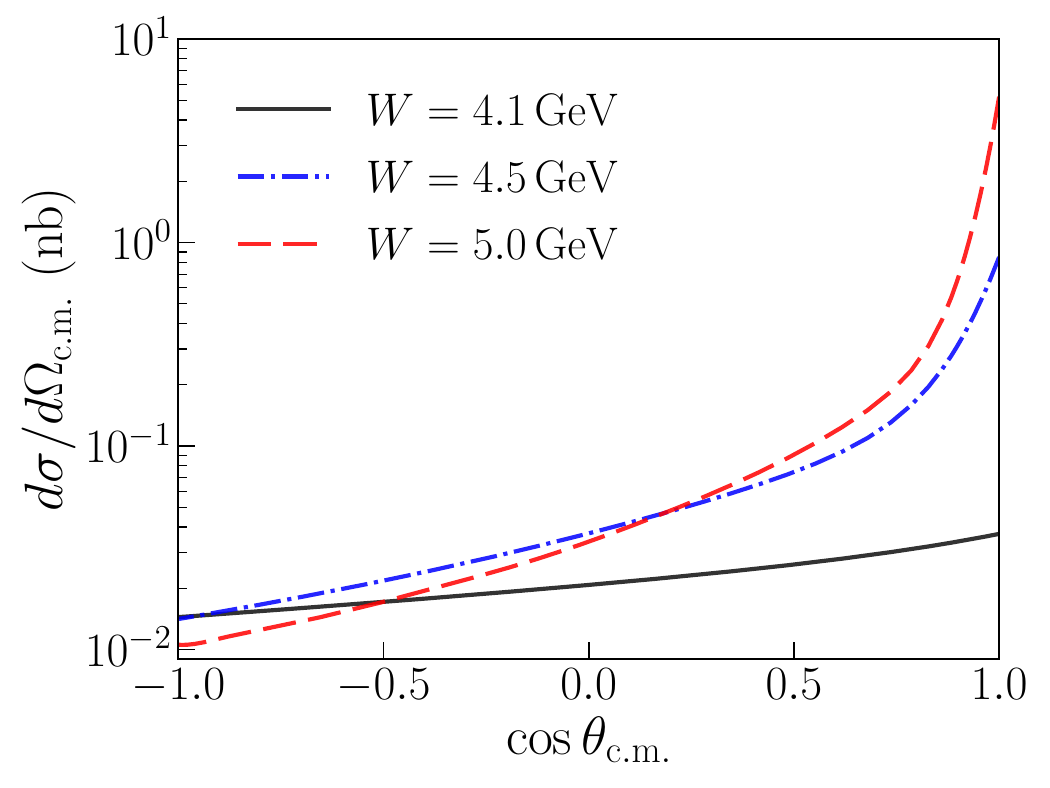}
\caption{The differential cross section of the photoproduction of $J/\psi$ from the nucleon in the Pom-pot model. The angles $\Omega_{\rm c.m.}$ and $\theta_{\rm c.m.}$ are defined in the center-of-mass frame with the photon moving along the $z$-direction.}
\label{fig:dsmodel}
\end{figure}

A feature of the VMD model using the lattice QCD potential is that the total cross section decreases at high energy, {\it i.e.} large $W$, where the data are orders of magnitude above the model prediction. A possible solution to this problem is to extend the VMD model by including the pomeron-exchange, which dominates the cross section at high energies. Although the $J/\psi-N$ potential at short distances, $r<0.4\,\rm fm$, is difficult to be quantified from the lattice QCD calculation~\cite{Kawanai:2010ev}, the long-distance region, which accounts for the nonperturbative interaction, can be well extracted. The effect of the less constrained short-distance region is discussed in Ref.~\cite{Lee:2020iuo}. To develop a model which contains both nonperturbative and perturbative mechanisms, one can formally add a perturbative potential term $V_{\rm pQCD}(r)$ into the Hamiltonian. The photoproduction amplitude will then obtain a term from $V_{\rm pQCD}(r)$ in addition to the amplitude in the original VMD model. Due to the absence of perturbative potential, one can further replace the amplitude from the $V_{\rm pQCD}(r)$ term by the amplitude from the pomeron-exchange model~\cite{Oh:2002rb,Wu:2012wta}, which is developed based on the perturbative analysis of Donnachie and Landshoff~\cite{Donnachie:1984xq}. 

By fitting the JLab data~\cite{Ali:2019lzf}, it was also observed that a rescaling factor $F^{\rm off}=0.4(0.7)$ to the VMD coupling $1/f_{J/\psi}$ was necessary to describe the near-threshold region while fit to high energy region is also improved. The introduction of such a correction factor can be interpreted as the scale dependence of the coupling and is consistent with the study of $\rho$ photoproduction~\cite{Donnachie:1994zb}. The real photon considered here has invariant mass square $q^2=0$, which is very different from the decay process $J/\psi \to \gamma \to e^+e^-$ with $q^2 = M_{\tiny J/\psi}^2 \sim 9\,\rm GeV^2$, conventionally used to determine the VMD coupling. Then the resulting Pom-pot model can describe data from near threshold up to $W=300\,\rm GeV$~\cite{Lee:2022ymp}. 

In Fig.~\ref{fig:dsmodel}, we show the differential cross section of the photoproduction of $J/\psi$ from the nucleon at several $W$ values near the threshold. More comprehensive comparison between the model predictions and experimental data, as well as the comparison among different models, can be found in Ref.~\cite{Lee:2022ymp}.

\subsection{Simulation of $J/\psi$ production from a deuterium target}

In this subsection, we describe the approach to simulate the photoproduction and the electroproduction of $J/\psi$ from a deuterium target, using the differential cross section of the photoproduction from the nucleon as a model input.

As a baseline estimation, we only consider the quasi-free mechanism of the production from the deuteron, and final-state interaction effects are neglected. In the quasi-free mechanism, subthreshold production is possible because of the Fermi motion of the nucleons in the deuteron nucleus. Here we use a two-pole parametrization~\cite{Tiburzi:2000je} of the deuteron wave function,
\begin{align}
 \widetilde{\Psi}({\bf p}) = \frac{1}{\sqrt{c}} \left[\frac{1}{{\bf p}^2 + a^2} - \frac{1}{{\bf p}^2 + b^2} \right],
 \label{eq:two-pole-wf}
\end{align}
where the parameters $a=0.0456\,\rm GeV$ is determined by the position of the nucleon pole, $b=0.2719\,\rm GeV$ is determined empirically from the average deuteron size, and $c$ is a normalization factor~\cite{Strikman:2017koc}. The momentum distribution of the nucleon in the deuteron is then given by the square of the wave function, $|\widetilde{\Psi}({\bf p})|^2$. As shown in Ref.~\cite{Strikman:2017koc}, the two-pole parametrization provides an excellent approximation to the realistic wave function in a wide range of momentum. In comparison with the solution from the AV18 NN potential~\cite{Wiringa:1994wb}, the two-pole parametrization agrees well with the realistic momentum distribution up to $|{\bf p}|\sim 0.3\,\rm GeV/c$, but under-estimates the density at high momentum region because of the absence of the $D$-wave component. Since this deviation will only result in possible under-estimates of the signal rate for the subthreshold and near-threshold productions, it will not affect the conclusion of the feasibility study in this paper.

As the simplest nucleus, the deuteron only contains two nucleons, and thus the momenta of the two nucleons is always back-to-back with the same magnitude in the deuteron rest frame. In the impulse approximation, one of the bound nucleons is struck by the photon to produce the $J/\psi$ as well as the recoil nucleon, while the spectator nucleon keeps its momentum as in the initial state. Then we can solve the energy of the struck nucleon,
\begin{align}
E_{\tiny N} = M_d - \sqrt{M_{\tiny N}^2 + {\bf p}^2},
\label{eq:off-shell-energy}
\end{align}
where $M_d$ and $M_{\tiny N}$ are the deuteron mass and the nucleon mass respectively. In the simulation, we randomly sample the nucleon momentum according to the distribution given by the wave function~\eqref{eq:two-pole-wf} and then calculate its off-shell energy from Eq.~\eqref{eq:off-shell-energy}. The $J/\psi$ events is generated via the subprocess $\gamma + ``p" \to J/\psi + p$ using the differential cross section from the Pom-pot model~\cite{Lee:2020iuo}. For the detection reason, we only consider the production from the proton inside the deuteron.

For the photoproduction, the target itself serves as the radiator. In the simulation, we generate photons according to an approximate distribution of the Bremsstrahlung spectrum~\cite{Zyla:2020zbs},
\begin{align}
\frac{{\rm d} N_{\gamma}}{{\rm d} E_{\gamma}} = \frac{d}{X_0} \frac{1}{E_\gamma}\left(\frac{4}{3} - \frac{4}{3}y + y^2\right),
\end{align}
where $E_\gamma$ is the photon energy, $y$ is the energy ratio between the Bremsstrahlung photon and the electron beam, $d$ is the target length, and $X_0$ is the radiation length of the target. The direction of the Bremsstrahlung photons is assumed along the beam direction. For liquid deuterium target, $X_0=769.1\,\rm cm$. In this study, we adopt $8.5\,\rm GeV$ electron beam incident on a 15-cm liquid deuterium target, corresponding to $d/X_0 \approx 1.95\%$. Since the target is utilized as the radiator, the effective target length for the photoproduction is half of the target length.

For the electroproduction, we consider the $J/\psi$ production from the subprocess $\gamma^* + ``p" \to J/\psi + p$, where $\gamma^*$ is the virtual photon. At the leading order of $\alpha$, the electromagnetic fine structure constant, we can relate the electroproduction cross section to the photoproduction cross section as
\begin{align}
 \frac{E' {\rm d}\sigma_{\tiny eN\to e' J/\psi N}}{{\rm d}^3 {\bf l}' d\Omega_{\rm c.m.}}
 &= \frac{\alpha}{2\pi^2 (1-\epsilon) Q^2}
 \frac{\sqrt{(q\cdot p)^2 - q^2 p^2}}{\sqrt{(l\cdot p)^2-l^2 p^2}}
 \nonumber\\
 &\times \frac{d\sigma_{\tiny \gamma^* N\to J/\psi N}}{d\Omega_{\rm c.m.}},
\end{align}
where $l$, $l'=(E',{\bf l}')$, $q=l-l'$, and $p$ are the four-momenta of the incoming electron, the scattered electron, the virtual photon, and the initial nucleon respectively. $Q^2=-q^2$ is the virtuality of the photon, and $\epsilon$ is the ratio between the timelike (also referred to as longitudinal) polarized and transverse polarized photons. This reaction is dominated by the low-$Q^2$ region, in which case it is referred to as quasi-real photoproduction.

In this study, both the photoproduction and the electroproduction processes are simulated. As the cross section of subthreshold production is small, we do not require the detection of the scattered electron to have more statistics of $J/\psi$ events. In such case, the events from the real and quasi-real photoproductions are indistinguishable. 

\begin{figure}[htp]
\centering
\includegraphics[width=0.8\columnwidth]{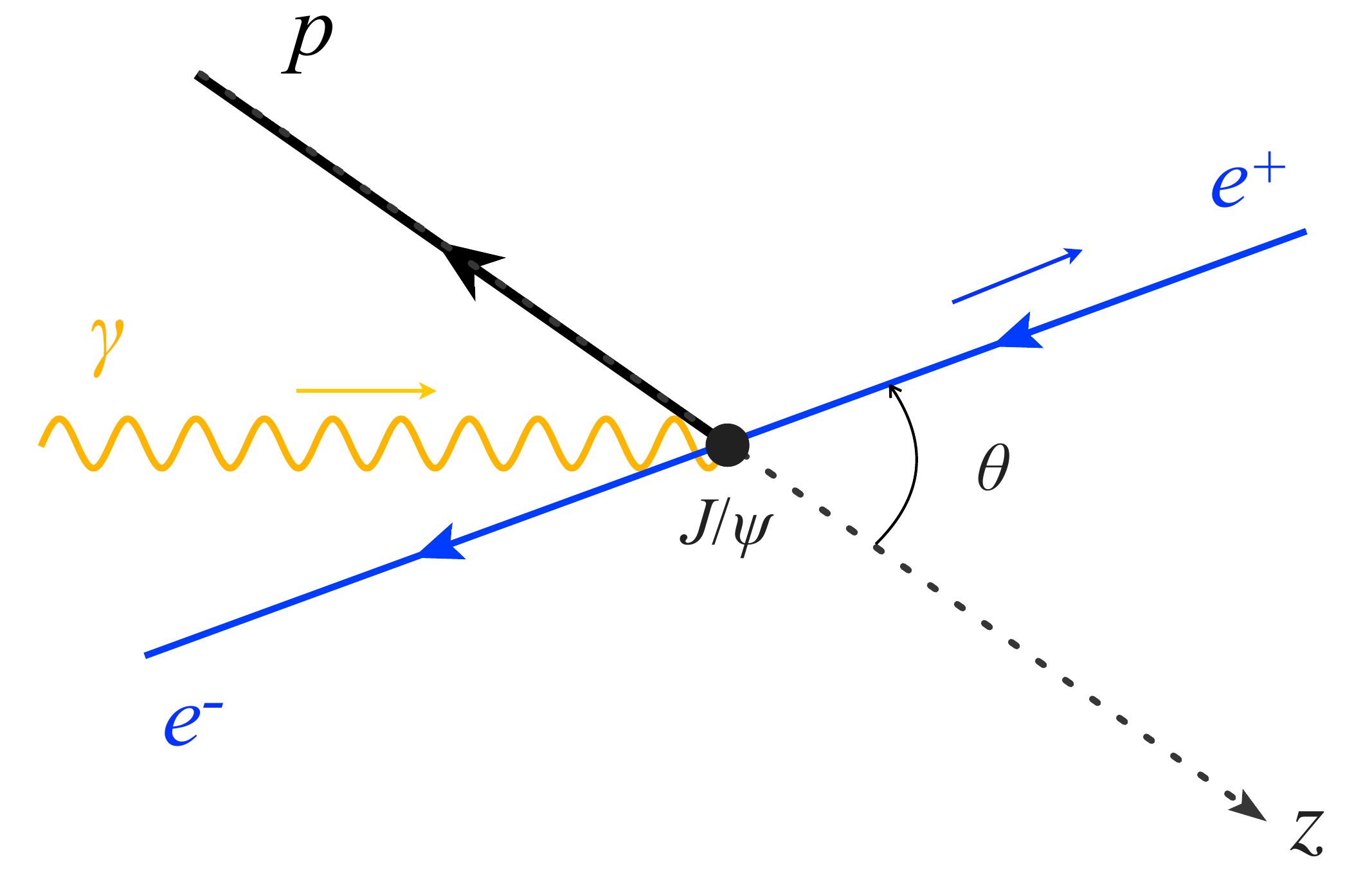}
\caption{The definition of the $\theta$ angle describing the decay angular distribution in the $J/\psi$ rest frame in Eq.~\eqref{eq:decaydistribution}.}
\label{fig:decayframe}
\end{figure}

The $J/\psi$ mesons are reconstructed from the decay channel $J/\psi\to e^+ e^-$. The $e^+e^-$ pair from the $J/\psi$ decay is assumed to follow the distribution~\cite{Schilling:1973ag},
\begin{align}
 f(\cos\theta) = \frac{3}{8\pi} [ 1 - \rho + (3\rho - 1) \cos^2\theta],
 \label{eq:decaydistribution}
\end{align}
where
\begin{align}
\rho &= \frac{\epsilon R}{1 + \epsilon R},
\end{align}
and~\cite{Martynov:2001tn,Martynov:2002ez}
\begin{align}
R &= \left(\frac{\lambda M_{\tiny J/\psi}^2 + Q^2}{\lambda M_{\tiny J/\psi}^2}\right)^n - 1.
\end{align}
The parameters $\lambda = 2.164$ and $n = 2.131$ are determined phenomenologically~\cite{Fiore:2009xk}. As illustrated in Fig.~\ref{fig:decayframe}, the angle $\theta$ of the positron is defined in the $J/\psi$ rest frame with the recoil nucleon moving against the $z$-direction.

For the study of subthreshold production of $J/\psi$ mesons and the exploration of the $J/\psi-N$ interaction, we require a 3-fold simultaneous detection of $e^+$, $e^-$, and $p$ in the final state. The detection of the recoil proton is not only important for the reconstruction of the untagged photon energy, but also necessary for deriving the relative momentum of the $J/\psi-N$ system, which is essential to extract the $J/\psi-N$ interactions.

With only the detection of the momenta of $J/\psi$ and the proton, one cannot determine the photon energy from the energy-momentum conservation for a general nuclear target. However, for the deuteron, as the simplest nucleus, the on-shell condition of the undetected nucleon in the final state provides an additional relation, which allow us to solve the photon energy as
\begin{align}
 E_\gamma = \frac{1}{2}\frac{(E_{\tiny J/\psi} + E_{p} - M_{d})^2 - ({\bf p}_{\tiny J/\psi}+{\bf p}_p)^2 - M_n^2}{E_{\tiny J/\psi} + E_p - p_{\tiny J/\psi}^z - p_{p}^z - M_d},
 \label{eq:photonenergy}
\end{align}
where $M_n$ is the mass of the undetected neutron, $E_{\tiny J/\psi}$ and $E_p$ are the energies of the produced $J/\psi$ and the proton, and ${\bf p}_{\tiny J/\psi}$ and ${\bf p}_p$ are their momenta with $z$-components $p_{\tiny J/\psi}^z$ and $p_{p}^z$. This relation does not rely on the quasi-free mechanism or the impulse approximation, and it is a unique advantage of the deuteron target.

As demonstrated in Ref.~\cite{Wu:2013xma}, when the relative momentum of the $J/\psi-N$ system is small, the reaction is more sensitive to the $J/\psi-N$ potential. To select the events in such region, we introduce the momentum of $J/\psi$ in the $J/\psi-N$ c.m. frame, labeled as $\kappa_{_{J/\psi}}$, which can be calculated from the momenta of the $J/\psi$ and the proton as
\begin{align}
\kappa_{_{J/\psi}} = \frac{1}{2W}\sqrt{(W^2 - M_{\tiny J/\psi}^2 - M_p^2)^2 - 4M_{\tiny J/\psi}^2 M_p^2}.
\label{eq:kappa}
\end{align}

\subsection{Simulation results}

Following the procedure described above, we simulate the $J/\psi$ production events using a $8.5$-GeV electron beam with $1.25\,\rm\mu A$ current incident on a 15-cm liquid deuterium target, corresponding to the luminosity of $1.2\times10^{37}~{\rm nucleons}~{\rm cm}^{-2}\cdot s^{-1}$. The SoLID acceptance, as shown in Sect.~\ref{sec:setup}, is applied for final state particles. Since no neutron detection is designed in SoLID and the detection of protons requires the momentum greater than $0.2\,\rm GeV/c$, we do not consider the $J/\psi$ production from the neutron in the deuteron. To account for the detector resolution, the three-momenta of detected particles are randomly smeared with gaussian distributions according to the SoLID resolution described in Sect.~\ref{sec:setup}. For signal rate estimation, the results are scaled by a factor of $0.8$ as the overall detection efficiency. All results in this subsection are the sum of real and quasi-real photoproductions, since they are experimentally indistinguishable without the detection of the scattered electron.

\begin{figure}[htp]
    \centering
    \includegraphics[width=\columnwidth]{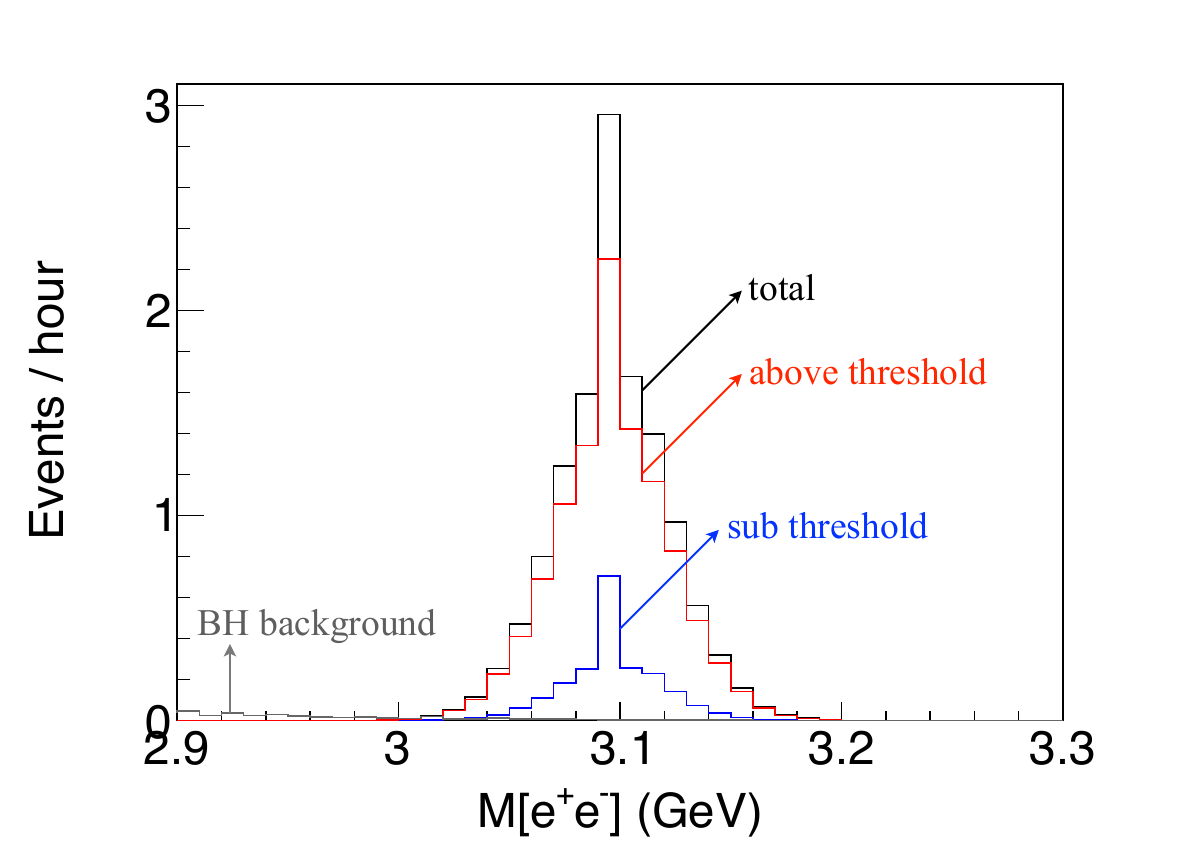}
    \caption{$J/\psi$ signal rate reconstructed from the $e^+e^-$ decay channel. The lower blue curve is the production below the threshold, the middle red curve is the production above the threshold, the the upper black curve is the sum of subthreshold and above-threshold productions. The gray curve is the background from the Bethe-Heitler process.} 
    \label{fig:mass}
\end{figure}

In Fig.~\ref{fig:mass}, we show the distribution of the invariant mass of the $e^+e^-$ pair from the $J/\psi$ decay. The width is dominated by the detector resolution since the intrinsic width of $J/\psi$ is negligible. To suppress the background while keeping the signal events as many as possible, we choose a $120\,\rm MeV$ window, {\it i.e.} $3.04\sim3.16\,\rm GeV$, around the $J/\psi$ mass. The signal rates and the total counts in 47 days are listed in Table~\ref{tab:rate}. The well-known Bethe-Heitler (BH) process contains the same final state particles as $J/\psi$ production. It could be the main background for the kinematics of study. We followed the recent BH calculation on deuteron~\cite{BH:2022} and carried out the same simulation. The resulting BH event rate is at sub-percent level comparing to the $J/\psi$ as shown in Fig.~\ref{fig:mass}.

\begin{table}[htp]
    \centering
    \caption{The event rates and the total counts in 47 days with the invariant mass of $e^+e^-$ within the $120\,\rm MeV$ window around the $J/\psi$ mass.}
    \begin{tabular}{lcc}
        \hline\hline
        &   ~~~~~~events/hour~~~~~~ & events in 47 days \\
        \hline
         subthreshold & 2.1 & 2360 \\
         above threshold & 10.3 & 11621 \\
         \hline
         total & 12.4 & 13981 \\
         \hline\hline
    \end{tabular}
    \label{tab:rate}
\end{table}

\begin{figure}[htp]
    \centering
    \includegraphics[width=\columnwidth]{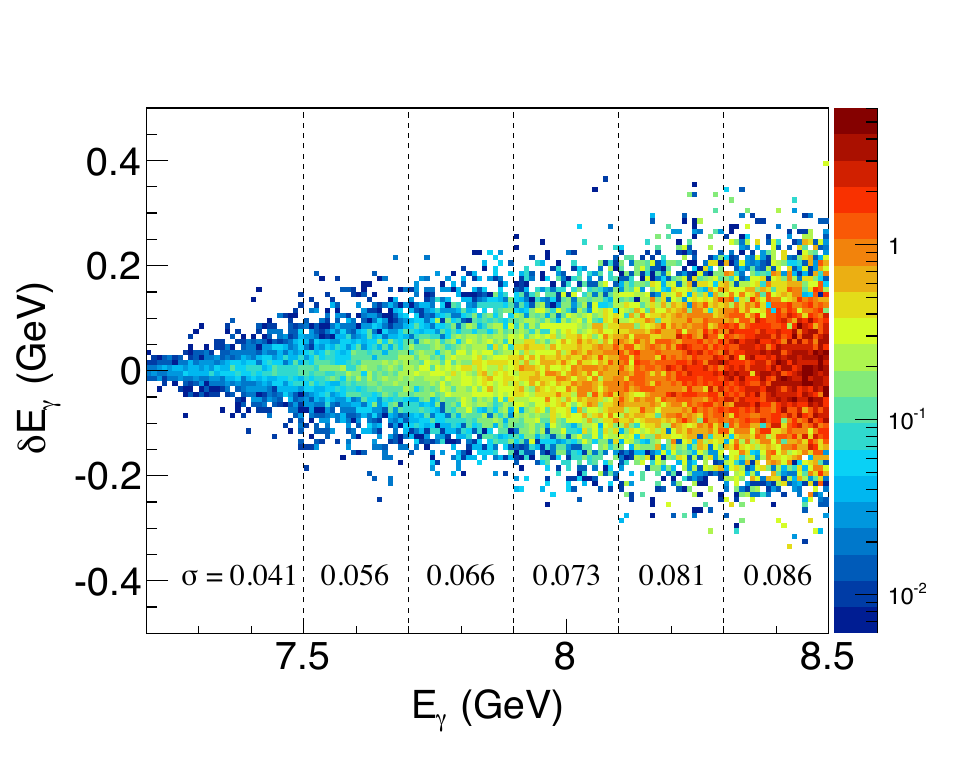}
    \caption{The smearing of the reconstructed photon energy due to the detector resolution. $E_{\gamma}$ is the photon energy recorded in the simulation. $\delta E_{\gamma}$ is the difference between the reconstructed photon energy and $E_\gamma$. The $\sigma$ labels the standard deviation of $\delta E_{\gamma}$ in each energy interval separated by vertical dashed lines. The color code represents the event rate distribution in the logarithmic scale.}
    \label{fig:gammares}
\end{figure}

\begin{figure}[htp]
    \centering
    \includegraphics[width=\columnwidth]{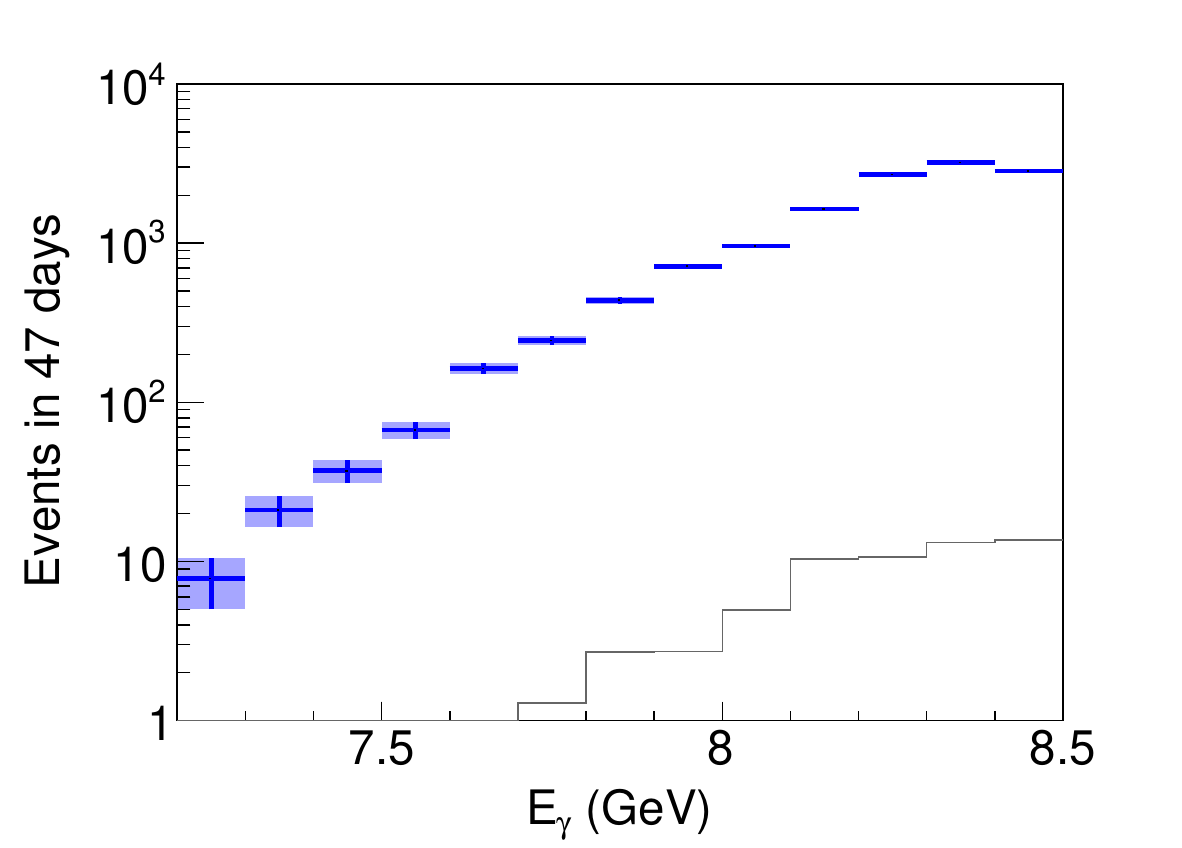}
    \caption{Projections in the photon energy. The horizontal width represents the binning and the vertical width represent the statistical uncertainty. The gray curve is the background from the Bethe-Heitler process.}
    \label{fig:gammaproj}
\end{figure}

\begin{figure}[htp]
    \centering
    \includegraphics[width=\columnwidth]{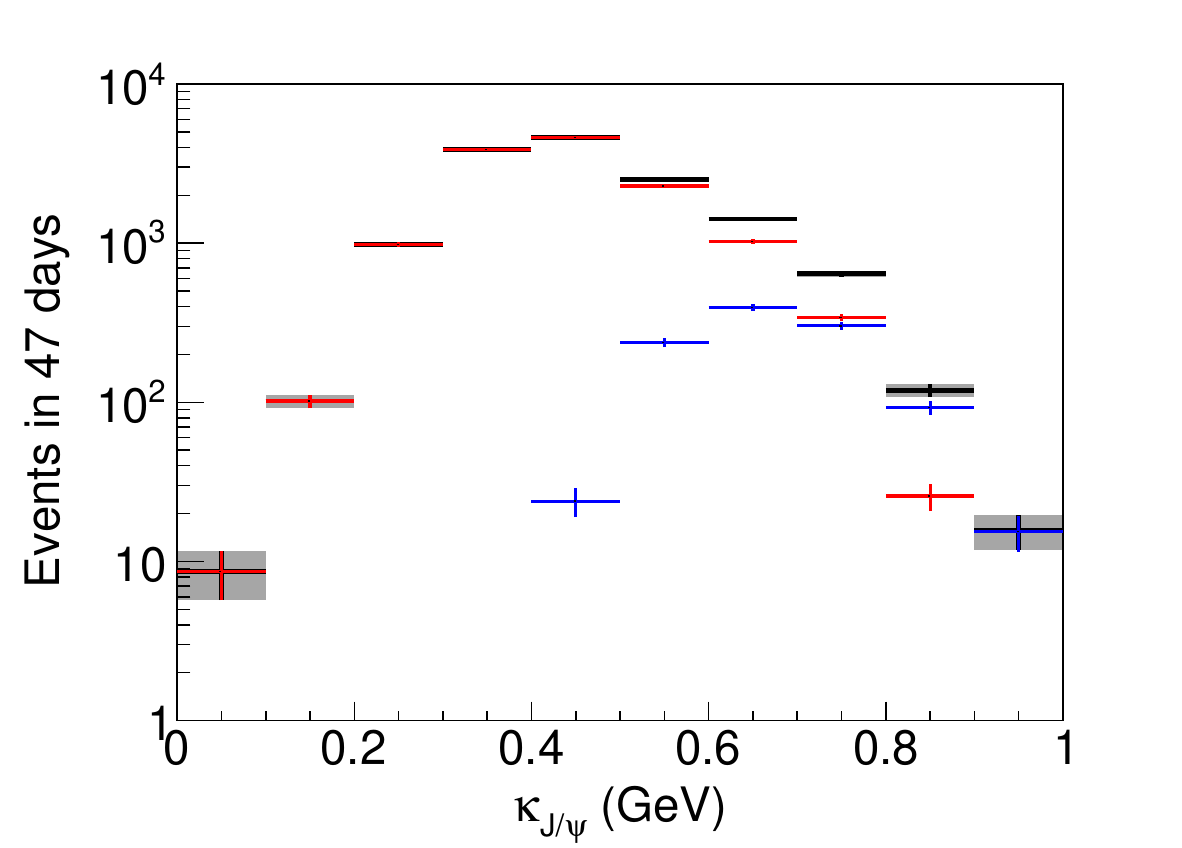}
    \caption{Projections in the $J/\psi$ momentum in the $J/\psi-N$ c.m. frame. The horizontal width represents the binning and the vertical width represents the statistical uncertainty. The blue points are the results with the proton momentum less than $1\,\rm GeV$, the red points are the results with the proton momentum greater than $1\,\rm GeV$, and the black points are the sum.}
    \label{fig:kappaproj}
\end{figure}

\begin{figure}[htp]
    \centering
    \includegraphics[width=\columnwidth]{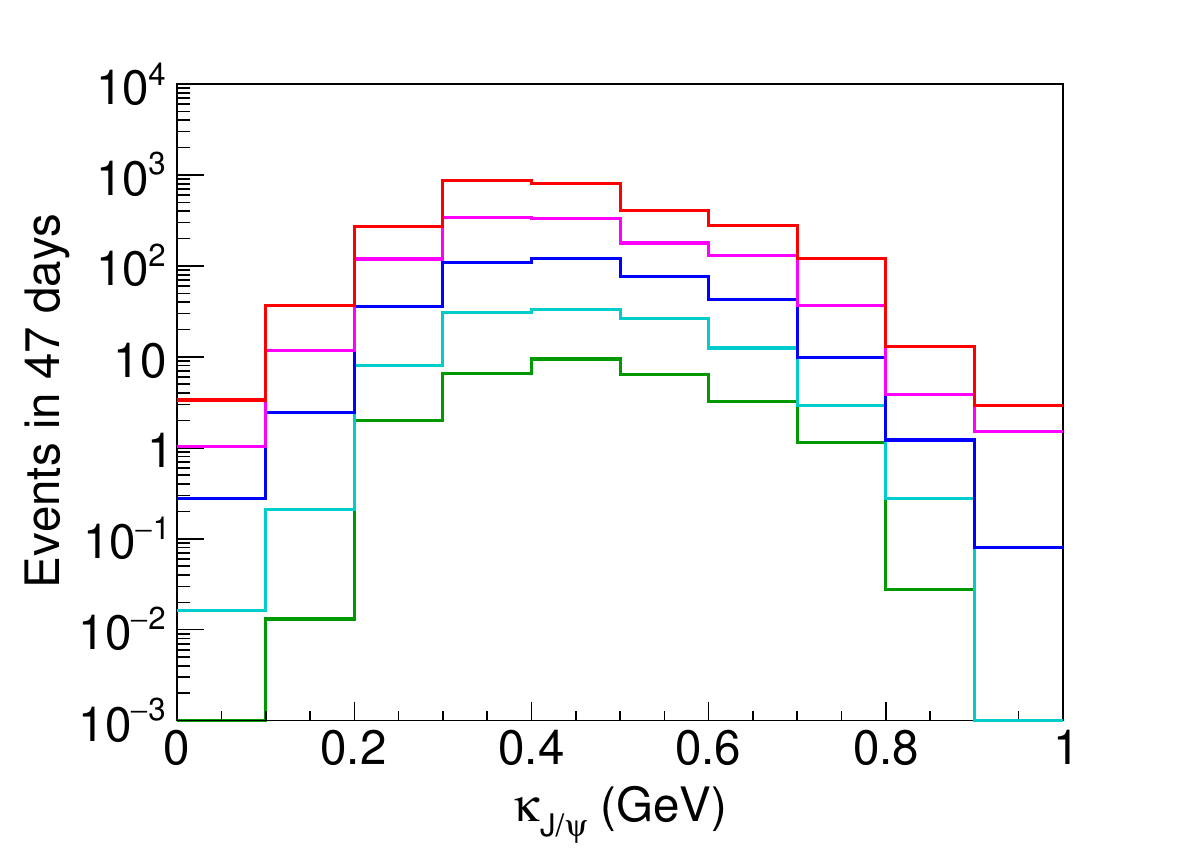}
    \caption{Distributions of the $J/\psi$ momentum in the $J/\psi-N$ c.m. frame from productions below the threshold. The curves from bottom to top correspond to the photon energy bins of $[7.2,7.4]\,\rm GeV$ (green), $[7.4,7.6]\,\rm GeV$ (cyan), $[7.6,7.8]\,\rm GeV$ (blue), $[7.8,8.0]\,\rm GeV$ (magenta), and $[8.0,8.2]\,\rm GeV$ (red).}
    \label{fig:kappagammadistri}
\end{figure}

Taking the unique advantage of the deuteron target, we can reconstruct the photon energy from Eq.~\eqref{eq:photonenergy} by a simultaneous detection of the recoil proton along with the two decay leptons from the $J/\psi$. The uncertainty of the determination of the photon energy depends on the detector resolution. In Fig.~\ref{fig:gammares}, we show the smearing of the reconstructed photon energy around the values recorded in the simulation. As labeled in the figure, the width of the smearing increases with the photon energy. The projection of the signal events in photon energy bins is shown in Fig.~\ref{fig:gammaproj}, where the statistical uncertainties are calculated from the square root of the number of events in the bins. The BH event rate is shown again at sub-percent level comparing to the $J/\psi$ for different photon energy.

The triple coincident detection of $e^+$, $e^-$, and $p$ also allows us to reconstruct the $J/\psi$ momentum in the $J/\psi-N$ c.m. frame via Eq.~\eqref{eq:kappa}. We observe from the simulation that the events with higher recoil proton momentum favor lower values of $\kappa_{_{J/\psi}}$. As demonstrated in Ref.~\cite{Wu:2013xma}, these events are more sensitive to the $J/\psi-N$ interaction. In Fig.~\ref{fig:kappaproj}, we show the projection in $\kappa_{_{J/\psi}}$ bins. The projections by selecting events with the proton momentum below and above $1.0~\rm GeV$ are also separately drawn. As shown in Fig.~\ref{fig:kappagammadistri}, the distributions of $\kappa_{_{J/\psi}}$ in several photon energy intervals, the expected statistics allows us to further bin the events in the photon energy, which will provide the opportunity of more detailed studies of the $J/\psi-N$ interaction.

\section{Discussion and Conclusions}
\label{sec:conclusions}

With the unique capability of SoLID in combining high luminosity and large acceptance, our studies have shown that subthreshold production of $J/\psi$ from the deuteron is experimentally feasible at JLab. The study on the deuteron also opens a new window for future studies of subthreshold production of $J/\psi$ from other nuclei and the data on the deuteron will serve as reference. Such studies will be important to search for $J/\psi$-nuclear bound states and nuclear short-range correlations. 
Further, the projected quasifree $J/\psi$ production data from the deuteron reported in this work, and future studies from other nuclei, near the production threshold of the proton can be compared with those from a free proton target to investigate any nuclear dependence in multi-gluon exchange.  Such studies can also help answer the question of whether there might be medium modification of the matter radius of the proton. 

$J/\psi$ meson near- and sub-threshold production from the deuteron is an important reaction to study non-perturbative QCD. Expanding from the proton, the production on the deuteron directly touches on many interesting QCD topics such as van der Waals interaction, multi-gluon exchange, bound states, short range correlations and gluonic gravitational form factors. While small cross section and multiple final state particle detection made this reaction a challenge to access in the past, the situation has changed since JLab entered the 12\,GeV era. 
Our study demonstrated this exciting measurement is now feasible through the next generation high luminosity and large acceptance detector SoLID. The high quality data would open a new door to allow many advanced QCD studies with this unique reaction.

\section{Acknowledgments}
\label{sec:ack}

We thank Dr. Tsung-Shung Harry Lee for his support with the production model and useful discussion. This article is based upon work supported by the
U.S. Department of Energy, Office of Science, Office of Nuclear
Physics under contract DE-FG02-03ER41231. 
It is also supported in part by the National Natural Science Foundation of China under grant No. 12175117, No. 12321005, and No. 20221017-1, and by the Shandong Provincial Natural Science Foundation under contract ZFJH202303.


\begin{thebibliography}{}
\bibitem{Zweig:1981pd}
G.~Zweig,
An SU(3) model for strong interaction symmetry and its breaking,
CERN-TH-401.
\end{thebibliography}
\end{document}